\documentclass[conference]{IEEEtran}
\usepackage{ifpdf, flushend,subfigure,tabularx}
\usepackage[left=1.62cm,right=1.62cm,top=1.9cm]{geometry}
\ifCLASSINFOpdf
  \usepackage[pdftex]{graphicx}
  \graphicspath{{../pdf/}{../jpeg/}}
  \DeclareGraphicsExtensions{.pdf,.jpeg,.png}
\else
  \usepackage[dvips]{graphicx}
  \graphicspath{{../eps/}}
  \DeclareGraphicsExtensions{.eps}
\fi
\usepackage{graphicx}
\usepackage{epstopdf}
\graphicspath{{../}}
\usepackage{multirow}
\usepackage{float}
\usepackage[cmex10]{amsmath}
\usepackage {amssymb}
\usepackage{array}
\usepackage{mdwmath}
\usepackage{mdwtab}
\usepackage{eqparbox}
\usepackage{nomencl}
\usepackage{cite}
\usepackage{algorithm}
\usepackage{algorithmic}
\usepackage{makeidx}
\usepackage{ifthen}
\usepackage{subfigure}
\usepackage{caption}
\usepackage{pdflscape}
\usepackage{hyperref}
\usepackage{xcolor}

\usepackage{amsmath} 
\usepackage{physics}
\usepackage{braket}

\usepackage{multirow}

\newcommand{\beq}{\begin{equation}}
\newcommand{\eeq}{\end{equation}}
\newcommand{\beqn}{\begin{eqnarray}}
\newcommand{\eeqn}{\end{eqnarray}}
\newcommand{\beqno}{\begin{eqnarray*}}
\newcommand{\eeqno}{\end{eqnarray*}}
\newcommand{\bma}{\begin{displaymath}}
\newcommand{\ema}{\end{displaymath}}
\newcommand{\bnu}{\begin{enumerate}}
\newcommand{\enu}{\end{enumerate}}
\newcommand{\bce}{\begin{center}}
\newcommand{\ece}{\end{center}}
\newcommand{\btb}{\begin{tabular}}
\newcommand{\etb}{\end{tabular}}

\begin{document}

\title{Quantum-based Distributed Algorithms for Edge Node Placement and Workload Allocation
}


\author{
    \IEEEauthorblockN{Duong~The~Do\IEEEauthorrefmark{2},   Ni Trieu\IEEEauthorrefmark{2},  and Duong Tung Nguyen\IEEEauthorrefmark{2}} \\
    \IEEEauthorblockA{\IEEEauthorrefmark{2}Arizona State University, Tempe, AZ 85281, USA,
    \{duongdo, nitrieu, duongnt\}@asu.edu} \\
}



\maketitle

\begin{abstract}
Edge computing is a promising technology that offers a superior user experience and enables various innovative Internet of Things  applications. In this paper, we present a mixed-integer linear programming (MILP) model for  optimal edge server placement and workload allocation, which is known to be  NP-hard. To this end, we  explore the possibility of addressing this computationally challenging problem using quantum computing.  However, existing quantum solvers are limited to solving unconstrained binary programming problems. To overcome this obstacle, we propose a hybrid quantum-classical solution that decomposes the original problem into a quadratic unconstrained binary optimization (QUBO) problem and a linear program (LP) subproblem. The QUBO problem can be solved by a quantum solver, while the LP subproblem can be solved using traditional LP solvers. Our  numerical experiments demonstrate  the practicality of leveraging quantum supremacy to solve complex optimization problems in edge computing.
\end{abstract}

\printnomenclature

\section{Introduction}
\label{Sec:Intro}
Edge computing (EC) expands upon traditional cloud computing by bringing cloud resources closer to end-users, things, and sensors. Consequently, this reduces latency and transmission costs, improves application performance, and ultimately enhances the user experience  \cite{b3, duongTON, duongTCC,ella23}. Without EC, all service requests are handled either locally or at remote cloud servers, which imposes a significant burden on  energy and bandwidth consumption. With EC, by utilizing computing resources at geographically distributed ENs, service providers can route user traffic to nearby ENs  more securely and efficiently \cite{b5}, drastically reducing latency and bandwidth consumption. 

Despite the enormous potential, there are many open problems in EC. This paper focuses on the joint optimization of EN placement and workload allocation in EC, with the objective of enhancing the user experience while minimizing cost. The goal of the EC platform is to minimize the total EN placement costs as well as  delay and unmet demand penalties. This problem is known to be NP-hard. To address this challenge, we formulate this problem as an MILP and propose to leverage quantum supremacy to solve this computationally challenging problem.

Quantum computing (QC) has become a focal point for the research community in recent years. The application of quantum mechanics to computing has resulted in the introduction of various quantum-based algorithms for solving complex problems in computer science. Recent advances in QC have enabled us to solve computational problems much faster than classical machines \cite{b6, b7}. Many computationally difficult combinatorial optimization problems, such as Max-Cut, Graph Coloring, and Traveling Salesman, can be mapped to Ising Hamiltonians whose ground states provide optimal solutions \cite{b9, b10, b11}. Current quantum computers have over 100 qubits and are capable of performing heavy computational tasks. Leading companies, such as Intel, IBM, and Google, are at the forefront of developing quantum computers based on superconducting technology \cite{b12}, while North American Rigetti is making significant efforts to build a fully functional quantum computer based on Josephson junctions.  American IonQ and Austrian AQT are using trapped ions to construct quantum computers \cite{b13}. Xanadu, a company specializing in photonic quantum computing, offers cloud access to its quantum computer via the D-Wave platform, a leader in quantum annealing \cite{b14, b15, b16}. 

The use of quantum processors has resulted in significant advancements in mathematical foundations and algorithm designs. Therefore, cloud/edge computing and related industries, including smart cities and manufacturing automation, can greatly benefit from the progress made in QC. Various attempts are underway to extend quantum  algorithms to NP-hard problems, utilizing the qubits' binary variable representation \cite{Farhi, b18}. However, these algorithms are typically heuristic. 
Therefore,  it is crucial to evaluate the practicality of using QC to optimize edge networks by analyzing the performance of these algorithms in edge network optimization applications.

This paper presents a novel approach to solving the joint EN placement and workload allocation problem using QC in a distributed manner. To the best of our knowledge, this is the first attempt to leverage  QC for this purpose. Specifically, we employ a hybrid quantum-classical approach that combines the alternating direction method of multipliers (ADMM) algorithm \cite{admm} with off-the-shelf quantum solvers. However, existing quantum solvers are limited to solving unconstrained binary programming problems, while the formulated MILP is a sophisticated constrained optimization problem. To address this limitation, we decompose the original  problem into a quadratic unconstrained binary optimization (QUBO) problem and a linear program (LP) subproblem. The QUBO problem can be solved using a quantum solver, such as QAOA, while the LP subproblem can be solved using traditional LP solvers. We utilize Qiskit, an SDK for performing quantum computations from IBM \cite{Qiskit}, to solve QUBO problems during the implementation of the ADMM-based distributed algorithm.

The remainder of the paper is organized as follows. Section II describes the system  model and problem formulation. Section III introduces the basics of quantum physics and  the state-of-the-art tools for quantum optimization followed by the  solution approach. Numerical results are shown in Section IV. Finally, conclusions are presented in Section V.

\begin{table}[ht] 
\centering
\caption{List of notations and abbreviations.}
\begin{tabular}{|c|l|}
\hline
Notation   & Description\\
\hline	
EN, AP & Edge Node, Access Point\\
\hline
$m$, $M$, $\mathcal{M}$ & Index, number, and set of areas (APs)\\
\hline
$n$, $N$, $\mathcal{N}$ & Index, number, and set of candidate EN locations\\
\hline
$B$ & Budget of the platform \\
\hline
$h_n$ & EN placement cost at location $n$\\
\hline	
$C_n$ & Resource computing capacity of EN $n$\\
\hline
$\lambda_m$ & Computing resource demand in area $m$\\
\hline
$u_m$ & Amount of unmet demand in area $m$\\
\hline
$\rho_m$ & Penalty for unmet demand in area $m$\\
\hline
$\beta$ & Delay penalty parameter\\
\hline
$d_{m,n}$ & Network link delay between area $m$ and EN $n$\\
\hline
$z_n$ & $\left\{0, 1\right\}$, $"1"$ if an EN is installed at location $n$\\
\hline
$x_{m,n}$ & Amount of workload from area $m$ to EN $n$\\
\hline
\end{tabular} \label{notation}
\end{table}

\section{System Model and Problem Formulation}
\label{Sec:ProblemFormulation}

\subsection{System Model}
\label{SubSec:Model}
\begin{figure}[ht!]
\centering
    \includegraphics[width=0.425\textwidth,height=0.2\textheight]{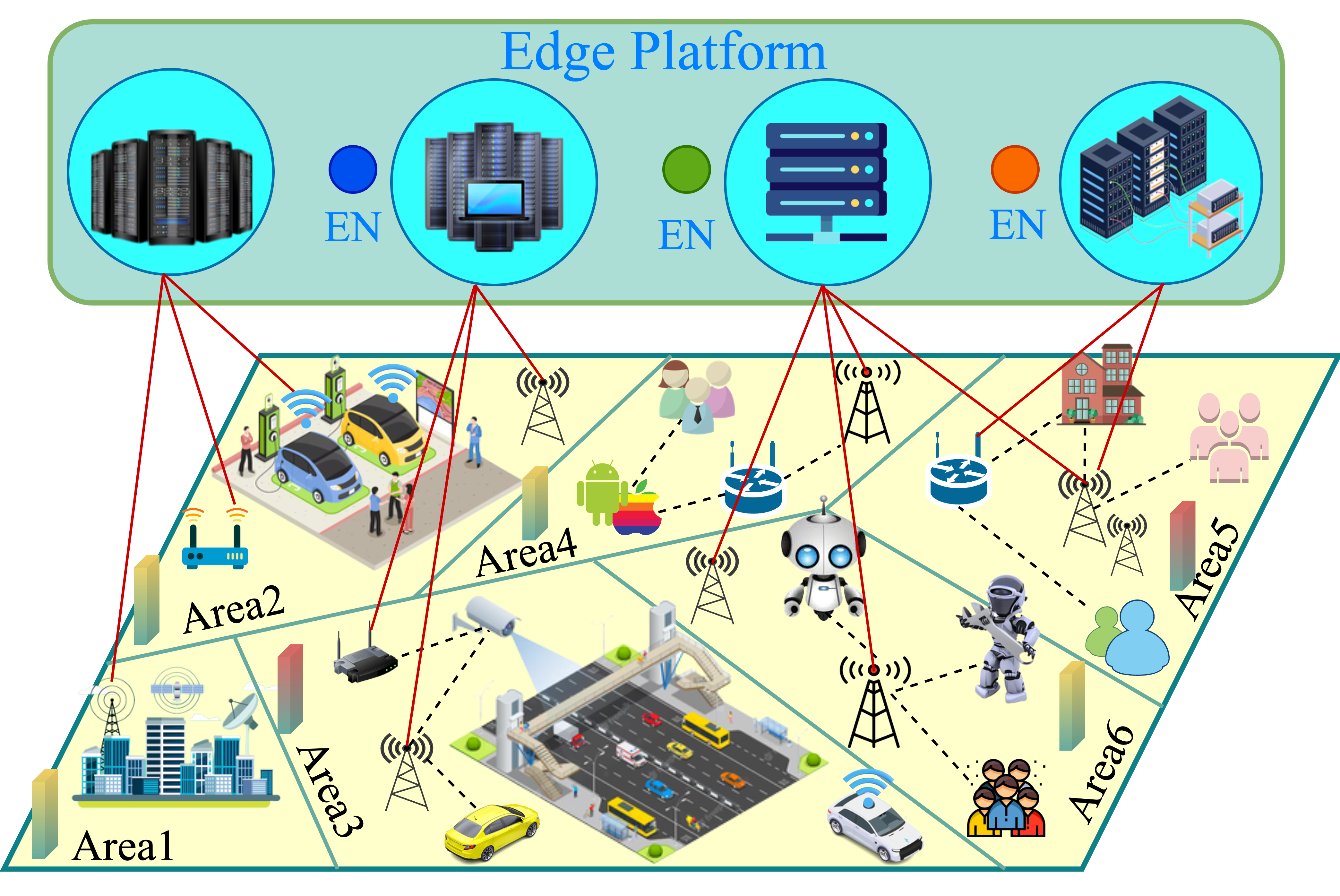}
	\caption{System model.}
\label{fig:model}
\end{figure}

The emerging EC  paradigm  brings computing and storage resources closer to end-users. 
 Typically, data and requests from users in each area are aggregated at an access  point (e.g., switches, routers, base stations) before being routed to edge nodes (ENs) or the cloud for further processing and analysis. We consider an EC platform that consists of numerous geographically distributed heterogeneous ENs  with different sizes and configurations. The system model, depicted in Fig.~ \ref{fig:model}, includes a set $\mathcal{N} = \left\{ 1, \cdots, N \right\}$ of $N$ potential candidate locations, managed by the platform for EN installation, as well as a set $\mathcal{M} = \left\{ 1, \cdots, M \right\}$ of $M$  areas served by the platform, each  represented by an access point (AP). The EN and AP indices are denoted by $n$ and $m$, respectively, and each EN may consist of one or several edge servers.

For simplicity, we consider only the computing resource and the capacity of EN $n \in \mathcal{N}$ is denoted by $C_n$. Also, define $\lambda_m \geq 0$ as the total resource demand (i.e., workload) of users in area $m \in \mathcal{M}$. Since IoT devices generate tremendous workloads demanding different IoT services, edge servers host IoT applications of different types to serve them. To reduce the delay between the users and the ENs, the demand from each area should ideally be served by its closest EN. However, the capacity of each EN is limited. Therefore, it is crucial for the platform to optimize workload allocation decisions, considering edge resource constraints as well as diverse locations of the ENs in order to ensure high service quality while lowering costs. Let $x_{m,n} \in \mathbb{R}$ be the amount of workload at the area $m \in \mathcal{M}$ assigned to the EN $n \in \mathcal{N}$. To simplify the notation, define $\boldsymbol{x}_m := \left[ x_{m,1}, x_{m,2}, \ldots, x_{m,N} \right]^T $, and $\boldsymbol{x} := \left[ \boldsymbol{x}_1^T, \boldsymbol{x}_2^T, \ldots, \boldsymbol{x}_M^T \right]^T$.

As the computing resources of the ENs are limited, they may not be able to fully satisfy the demand, resulting in dropped requests. This can negatively impact user experience, and thus the platform should heavily penalize unmet demand. To ensure user satisfaction, the problem of unmet demand must be considered. We denote the amount of unmet demand and the penalty for each unit of unmet demand in the area $m \in \mathcal{M}$ by $u_m$ and $\rho_m$, respectively. 
Furthermore, low latency is a critical requirement for delay-sensitive applications like remote robotics, augmented/virtual reality, and autonomous driving. Since network delay might cause serious QoS degradation, the EC platform must take delay into account while making optimal operations and planning decisions. 

In our model, we consider a graph $\mathbb{G}\left(\mathcal{V}, \mathcal{L}\right)$, where $\mathcal{V}$ is the set of nodes involving $M$ areas and $N$ ENs,  and $\mathcal{L} = \left\{\left(m, n\right) : m \in \mathcal{M}, n \in \mathcal{N}  \right\}$ is the set that represents the set of links connecting APs with ENs,  where $\left(m, n\right)$ indicates the link connecting area $m$ and EN $n$. We define a network delay between the AP $m \in \mathcal{M}$ and the EN $n \in \mathcal{N}$ as $d_{m,n} \in \mathbb{R}$. The goal of the platform is to minimize not only the total EN placement/installation cost but also the overall network delay and the amount of unmet demand. The main notations in this paper are summarized in Table \ref{notation}.

\vspace{-0.1in}
\subsection{Problem Formulation}
\label{SubSec:Formulation}
\vspace{-0.1in}
In this section, we develop an MILP model to capture the EN placement and workload allocation problem. First, we describe different cost components in the objective function of the platform.

\subsubsection{Installation cost}
Let  $z_n \in \left \{ 0, 1 \right \}$ be a binary indicator which takes the value of $1$ if an EN is installed at location $n \in \mathcal{N}$ and $0$ otherwise. If the platform decides to install an EN at the location $n \in \mathcal{N}$, it incurs an EN placement cost of $h_n \geq 0$. The total EN placement cost is given as:
\begin{align}
\label{eq-cost:placement}
    & \mathcal{C^P} = \sum_{n \in \mathcal{N}} h_n \, z_n.
\end{align}

\subsubsection{Network delay cost} The delay cost between  area $m$ and  EN $n$ is proportional to the product of the amount of workload from  area $m$ to EN $n$ as well as the network delay between them. As a result, the overall network delay penalty cost can be expressed as follows:
\begin{align}
\label{eq-cost:delay}
    & \mathcal{C^D} = \beta \sum_{m \in \mathcal{M}} \sum_{n \in \mathcal{N}} d_{m,n} \, x_{m,n},
\end{align}
where $\beta \in \mathbb{R}$ is defined as a weighting factor that represents the delay penalty cost parameter to control the tradeoff between EN placement cost and the service quality.

\subsubsection{Unmet demand cost}
The penalty for unmet demand in each area $m \in \mathcal{M}$ is proportional to the amount of unmet demand and the penalty for each unit of unmet demand $\rho_m$. Hence, the total penalty  $\mathcal{C^U}$  for unmet demand over all the areas is given by:
\begin{align}
\label{eq-cost:undemand}
    & \mathcal{C^U} = \sum_{m \in \mathcal{M}} \rho_{m} \, u_{m},
\end{align}
in which $\rho_m \geq 0$ is the unmet demand penalty parameter that can be adjusted by the platform. A higher value of $\rho_m$ implies that  area $m$ is more prioritized. 

We are now ready to formulate the EN placement and workload allocation problem which aims to 
determine optimal locations for EN installation, while improving user experience by minimizing both the overall network delay and unmet demand. 
Mathematically, this joint EN placement and workload allocation problem can be formulated as follows:
\begin{subequations}
\label{mainprob}
\begin{align}
\label{eq:objective}
    & \! \mathop{\text{min}}_{\boldsymbol{z, x, u}} \sum_{n=1}^{N} \! h_n z_n + \beta \! \! \sum_{m = 1}^{M} \sum_{n = 1}^{N} \! d_{m,n} x_{m,n}  + \! \! \sum_{m=1}^{M} \! \rho_m u_m  \\
    & \; \text{s.t.} \quad \text{(C1)}: \sum_{n=1}^{N} h_{n} \, z_{n} \leq B, \\
    & \quad \quad \; \, \text{(C2)}: 0 \leq \sum_{m=1}^{M} x_{m,n} \leq C_n \, z_n, ~ \forall n \in \mathcal{N}, \\
    & \quad \quad \; \, \text{(C3)}: \sum_{n=1}^{N} x_{m,n} + u_m = \lambda_m, ~ \forall m \in \mathcal{M}, \\
    & \quad \quad \; \, \text{(C4)}: z_n \in \{0, 1\}, ~ \forall n \in \mathcal{N}, \\
    & \quad \quad \; \, \text{(C5)}: u_{m}, \, x_{m,n} \geq 0, ~ \forall m \in \mathcal{M}, \forall n \in \mathcal{N}.
\end{align}
\end{subequations}

Here,  constraint $\text{(C1)}$ indicates that the total expense for the planning decisions should not exceed the restricted investment budget $B$ of the platform, whereas  constraint $\text{(C2)}$ enforces that the total allocated computing resource required to serve the total amount of workload allocated to a deployed EN $n$ can not surpass its maximum capacity $C_n$. Furthermore,  $\text{(C2)}$ implies that workload cannot be served at a location without EN installation. Specifically, it can be seen that if $z_n = 0 $, then $x_{m,n} = 0$, $\forall m,n $. The workload allocation constraint $\text{(C3)}$ imposes that the workload from each area must be either served by some ENs or dropped (i.e., counted as the unmet demand $u_{m}$). Finally, constraints $\text{(C4)}$ and $\text{(C5)}$ indicate the feasible set of decision variables. It is worth noticing that our model can be easily extended to  include additional system and design constraints, such as multiple types of resources (e.g., RAM, CPU, bandwidth) and various system uncertainties \cite{duongiot}.

\section{Integrated Quantum ADMM Algorithm}
\label{Sec:Algorithm}
The MILP problem (\ref{mainprob}) is a challenging one to solve due to its large-scale nature. Hence, we propose to leverage QC to tackle this issue.  
In this section, we introduce a quantum-based distributed algorithm that utilizes a combination of the ADMM framework and quantum solver to iteratively solve the joint EN placement and workload allocation problem (\ref{mainprob}) in a distributed manner. We will provide a brief overview of QC, followed by a discussion on how we can leverage quantum combinatorial optimization algorithms to solve the distributed EN placement problem, using a combination of both  quantum machines and classical computers.

\subsection{Fundamental of Quantum Computing}
\label{SubSec:Fundamental_QC}
QC relies on the principles of quantum mechanics to store and manipulate information in computational systems. 
 In contrast to classical computing,  which represents information using Boolean bits represented by physical quantities with two states to represent logically either 0 or 1 (e.g., magnetization direction in a magnetic drive), quantum computers use quantum bits, or qubits, to represent information. Each qubit can exist in multiple states simultaneously, unlike the classical bit, and can be in a superposition of two basis states $\left( \ket{0}, \ket{1} \right)$. This enables more complex operations and computations to be performed on quantum systems. For instance, a single-qubit's state can be described as a coherent superposition of the basis states, expressed as a linear combination of the basis states with complex probability amplitudes. 
\begin{align}
\label{eq:objective}
    & \ket{\psi} = \alpha \ket{0} + \beta \ket{1} \in \mathbb{C}^2 ,
\end{align}
where coefficients $\alpha$ and $\beta$ are the complex numbers, called the probability amplitudes with $\left\| \alpha \right \|^2 + \left \| \beta \right \|^2 = 1$. This  means that if we  measure $\ket{\psi}$ to see whether it is in $\ket{0}$ or $\ket{1}$, then the state collapses immediately to either $\ket{0}$ with a real probability of $\left\| \alpha \right \|^2$ or $\ket{1}$ with a probability of $\left \| \beta \right \|^2$. Similarly, by using tensor products, the configurations of a two-qubit system are the four configurations $\left\{\ket{0} \otimes \ket{0}, \ket{0} \otimes \ket{1}, \ket{1} \otimes \ket{0}, \ket{1} \otimes \ket{1}\right\}$ together. This can be generalized for a higher dimensional single qubit, in a generic form, an arbitrary state of an $n$-qubit quantum system described by a $2^n$ dimensional complex vector space $\mathbb{C}^{2^n}$ can be expressed through a linear combination of possible configurations of the basis states by suitable complex weights \cite{Burgholzer}.
As a result, quantum computers can store and process exponentially more information than classical computers, as $n$ qubits can represent $2^n$ possible states, equivalent to $2^n$ bits of information. Taking advantage of the huge computation space, roughly speaking, superposition allows quantum computers to analyze far more possibilities than a classical one. It is expected that a polynomial time of a quantum machine is more powerful and can be superior to that of a classical computing machine in certain search algorithms, making QC immensely helpful in solving large and complex problems.

\begin{figure}[h!]
    \subfigure[$\text{ibmq-jakarta}$]{
        \includegraphics[width=0.125 \textwidth, height=0.1 \textheight]{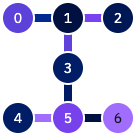}
	    \label{fig:IBM1} }
	    \hspace*{0.2em} 
		 \subfigure[$\text{ibmq-washington}$]{
	     \includegraphics[width=0.25 \textwidth, height=0.14 \textheight]{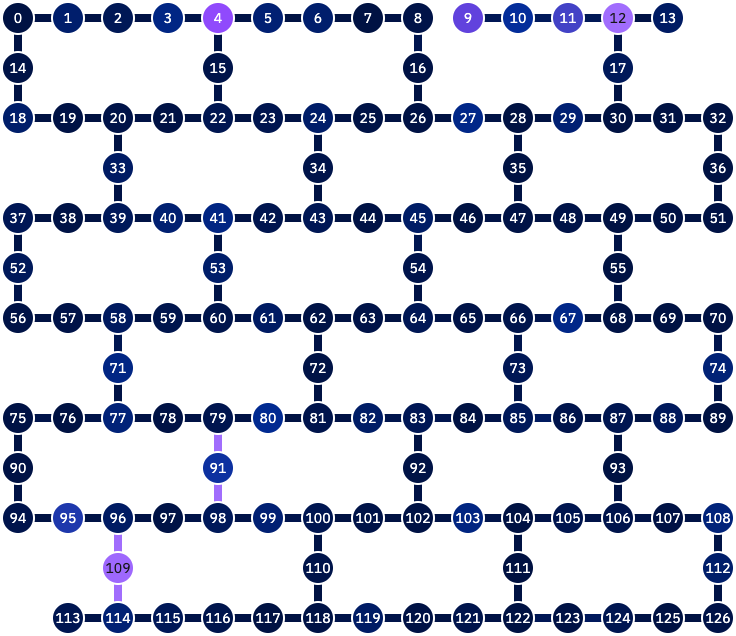}
	     \label{fig:IBM2}
	}  \vspace{-0.2cm}
	\caption{The structures of various quantum computers.}
\end{figure}

In the current era of QC, there is a growing interest in utilizing quantum solutions to address practical problems. Researchers are working to expedite the convergence, accuracy, and scalability of quantum algorithms for combinatorial optimization, which poses significant challenges for classical computers. To this end, various QC platforms, services, and development kits, such as Azure Quantum, AWS Quantum, D-Wave System, IonQ, Q\# and IBM Qiskit, have been developed, to accelerate the application of QC to combinatorial optimization. IBM occupies a unique position in the quantum computing landscape by offering its hardware and services to the general public, which includes multiple hardware options with varying degrees of computing capability, such as the number of qubits and coupling maps.  For instance, the coupling map for IBM Quantum Jakarta, a quantum computer with $7$ qubits available for free access, and IBM Quantum Washington, a quantum computer with $127$ qubits accessible through contracted access, are shown in Fig. \ref{fig:IBM1} and \ref{fig:IBM2}, respectively. Access to various quantum hardware empowers researchers to verify their computations against expected results, thereby enhancing overall research and algorithms.

Solving combinatorial optimization problems is a challenging task for any computer, including quantum computers. As a result, most existing quantum algorithms for solving these problems employ heuristics, which entail relaxing binary constraints by utilizing quantum states to produce approximate solutions  \cite{Farhi}.
Executing these heuristics on high-performance quantum hardware has the potential to enhance the accuracy of the resulting approximate solutions. Therefore, it is crucial to explore the effectiveness of quantum algorithms in solving combinatorial optimization problems, particularly those that are relevant to optimal operations and planning in  EC.

\subsection{Quantum Approximate Optimization Algorithm (QAOA)}
\label{SubSec:QAOA}
The quantum approximate optimization algorithm (QAOA) is a meta-heuristic for solving combinatorial optimization problems that can utilize gate-based quantum computers and possibly outperform purely classical heuristic algorithms. It shows great potential for demonstrating quantum advantage in the near future and is considered among the top contenders for noisy intermediate-scale quantum (NISQ) computation, primarily owing to the low number of qubits and gates generally needed \cite{Braine_2021}. The QAOA relies not only on the output of the quantum circuit but also on the classical optimizer that updates the circuit parameters to improve the solution of the optimization problem. Specially, this algorithm takes on combinatorial optimization problems, in particular problems that can be cast as searching for an optimal bitstring and aims to provide an approximation to the solution with the desired level of accuracy. For example, given a combinatorial optimization problem that is to find a value $x \in \left\{0,1\right\}^n$ that minimizes the value of the objective function $f(x)$ as $f: \left\{0,1\right\}^n \rightarrow \mathbb{R}$. The $n$ binary variables in the original bianry optimization are considered in a form of $n$-bit strings, $x_1x_2\cdots x_n$ and instead of minimizing $f(x)$, the QAOA converts the original problem into the searching for optimal decision variables in a form of bitstring by looking for the ground state (an eigenstate with the lowest associated eigenvalue) of an encoded cost Hamiltonian $\mathcal{H}_C$ of a quantum system, where $\mathcal{H}_C \ket{x} = f(x) \ket{x}$, as displayed in Fig. \ref{fig:QAOA}. The QAOA consists of the following steps:

\begin{figure}[ht!]
\centering
    \includegraphics[width=0.4\textwidth,height=0.12\textheight]{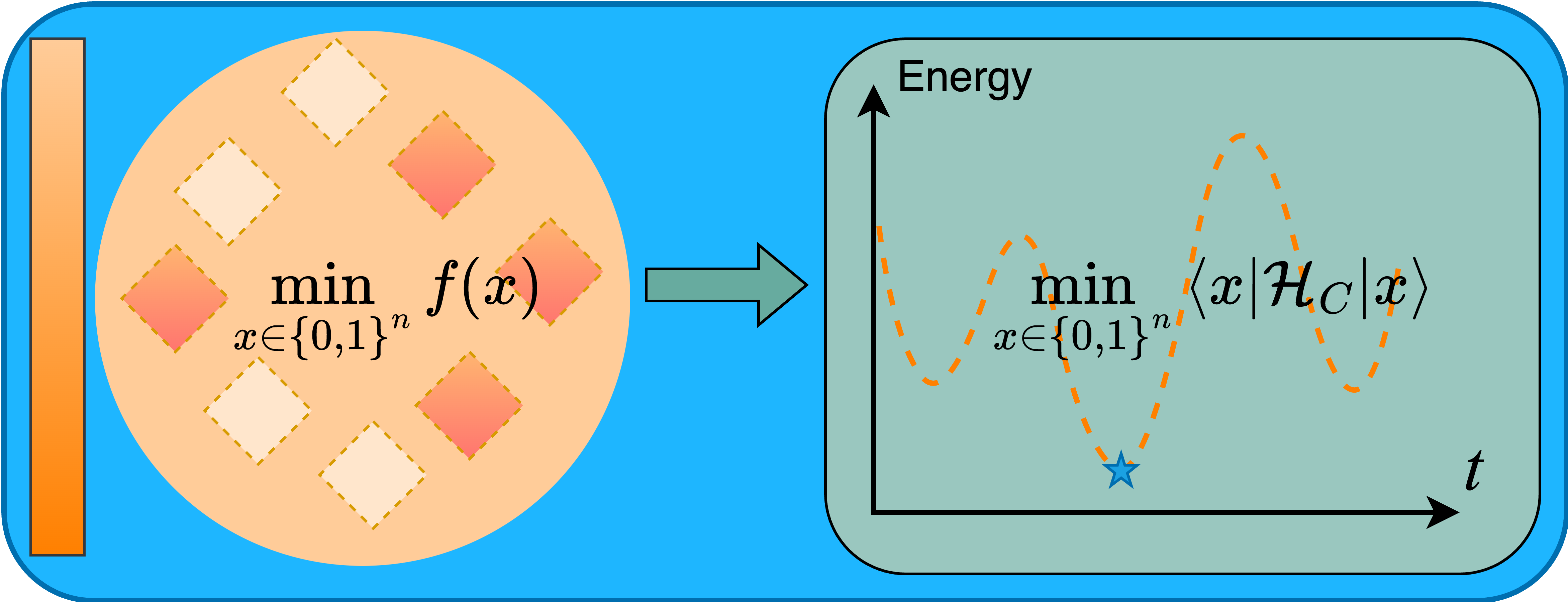}
	\caption{Illustration of encoding combinatorial optimization problems to QAOA.}
\label{fig:QAOA}
\end{figure} 

\begin{enumerate}
  \item Define a cost Hamiltonian $\mathcal{H}_{C}$ such that its ground state encodes the solution to the optimization problem and an adding Hamiltonian $\mathcal{H}_{A}$.
  \item Construct an Ansatz circuit by integrating unitary transformations \(e^{-i \alpha \mathcal{H}_{A}}\), \(e^{-i\beta \mathcal{H}_{C}}\) with a parameter selection of \(p\geq 1\), which determines the depth of the circuit. \[\mathbb{U}\left(\alpha,\beta\right) = \prod_{j=1}^p e^{-i\alpha_j\mathcal{H}_A} e^{-i\beta_j\mathcal{H}_C}.\] 
  \item Prepare an initial state, apply \(\mathbb{U}\left(\alpha,\beta\right)\), and use classical techniques to optimize the parameters. 
  \item Once the circuit has been optimized, measuring the output state provides approximate solutions to the optimization problem.
\end{enumerate}

\subsection{Quantum Based Distributed Solution Approach}
\label{SubSec:HQC_ADMM}
The EN placement and workload allocation problem described in equation (\ref{mainprob}) involves both continuous and integer variables, which makes it difficult to solve using traditional MILP solvers like GUROBI, MOSEK, CPLEX, and SCIP. These solvers are not scalable for large-scale problems, and their running time increases exponentially as the problem size increases. On the other hand, existing quantum solvers are limited to solving
unconstrained binary programming problems.

To this end, we propose to employ a hybrid quantum ADMM distributed algorithm \cite{admmqc} to efficiently solve the proposed problem (\ref{mainprob}). 
The hybrid quantum-classical ADMM-based (HQC-ADMM) distributed algorithm combines the ADMM framework, a well-known distributed algorithm for solving large-scale convex optimization problems, and quantum solvers to solve the MILP (\ref{mainprob}) on both quantum and classical machines. The ADMM framework decomposes a large-scale problem into smaller subproblems that can be solved in parallel by different agents in the system.
Various ADMM-based distributed algorithms have been proposed to tackle numerous real-world engineering applications. Recently, ADMM has been used to tackle combinatorial optimization problems as well \cite{admmqc}. While the HQC-ADMM algorithm for MILP (i.e., a non-convex optimization problem) is heuristic, it can produce an exact optimal solution under certain conditions. Indeed, as demonstrated in our numerical experiments, the HQC-ADMM solution is very close to the exact optimal solution produced by MILP solvers for our problem (\ref{mainprob}).
\begin{figure}[ht!]
\centering
    \includegraphics[width=0.3\textwidth,height=0.18\textheight]{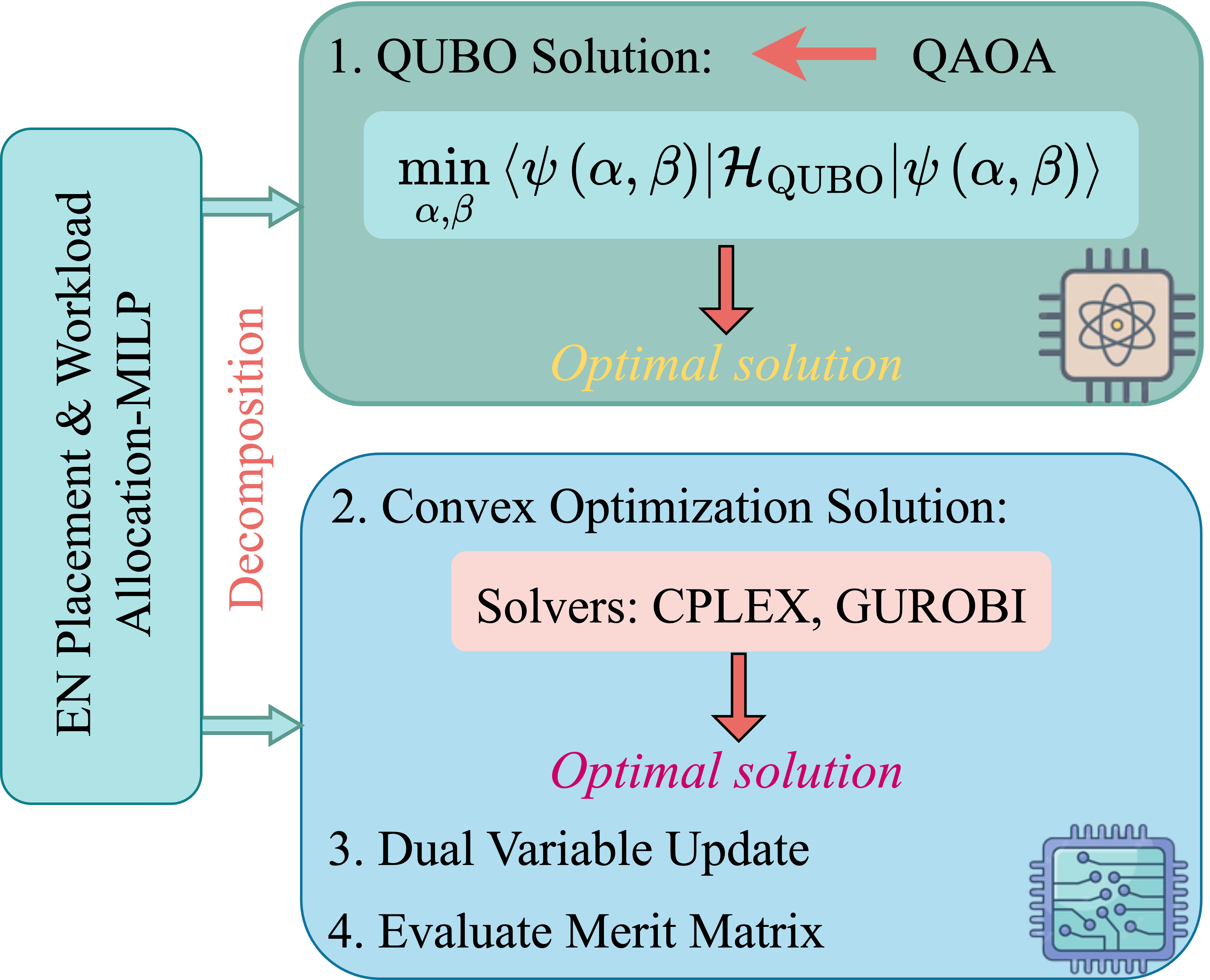}
	\caption{The HQC-ADMM-based distributed algorithm.}
\label{fig:ADMM}
\end{figure}  

This approach is based on the idea of exploiting the special structure of the problem, it partitions the original formulation into two smaller easier-to-solve problems that can be solved independently. Our approach  combines quantum and classical methods. We develop an iterative hybrid algorithm based on the ADMM procedure and  QAOA based on a QUBO formalism to solve the formulated EN placement and workload allocation problem. Specifically, the reformulated  QUBO problem is solved by  QAOA by relaxing binary by quantum states whereas the other linear convex problems are solved by conventional solvers such as CPLEX, GUROBI, SDPT3, and COBYLA, as displayed in Fig. \ref{fig:ADMM}. The EN placement decision is embedded into the QUBO problem while the workload allocation decision is captured by the LP subproblems.

The developed HQC-ADMM algorithm is implemented within a multi-block framework, which decomposes the original problem into smaller subproblems and solves them in parallel at each iteration. The algorithm iteratively solves an updated QUBO problem using a QUBO oracle or quantum devices and an updated convex problem using an off-the-shelf solver in each iteration until convergence.  However, since the results obtained by quantum computing are affected by the inexact outcome of the binary subproblem and noise generated by real quantum computers, ADMM cannot guarantee convergence to an optimal solution due to the non-smooth nature of the MILP problem.  The multi-block ADMM  framework is summarized in Fig. \ref{fig:ADMM}, which shows the proposed approach and implementation choices using QAOA as quantum QUBO solvers. Due to space limitations, we do not present the detailed algorithm here.

\section{Numerical Validation}
\label{Sec:NumericalResults}
In the experiments, following previous works \cite{Jia, Jia2, tara22}, a random scale-free edge network topology with 50 nodes is generated using the Barabasi-Albert model \cite{Albert}. The link  delays between adjacent nodes are randomly generated  in the range of $2 ms$ to $5 ms$ \cite{tara22}. The network  delay between any two nodes is the delay of the shortest path between them. The ENs are chosen randomly from the set of Amazon \textit{EC2 M5} instances \cite{EC2}. Additionally, the EN placement costs are randomly generated between  $\left[ 0.2, 0.25 \right]$. The budget of the SP is set to be $20$. The delay penalty parameter  is set to $10^{-4}$. The algorithm was implemented in Python on a machine with a $2.2$ GHz Intel Core i5 processor and 8GB of RAM. The simulations on quantum devices to solve the QUBOs were conducted  using Qiskit, an open-source SDK framework. Specifically, we utilized Qiskit version $0.15.0$, qiskitaqua version $0.6.1$, qiskit-terra version $0.10.$0, and qiskit-aer version $0.3.2$ \cite{Qiskit}. Additionally, IBM ILOG CPLEX $12.8$ was chosen as the classical optimization solver. 

\begin{figure}[h!]
    \subfigure[$N=3, M=5$]{
        \includegraphics[width=0.245 \textwidth, height=0.135 \textheight]{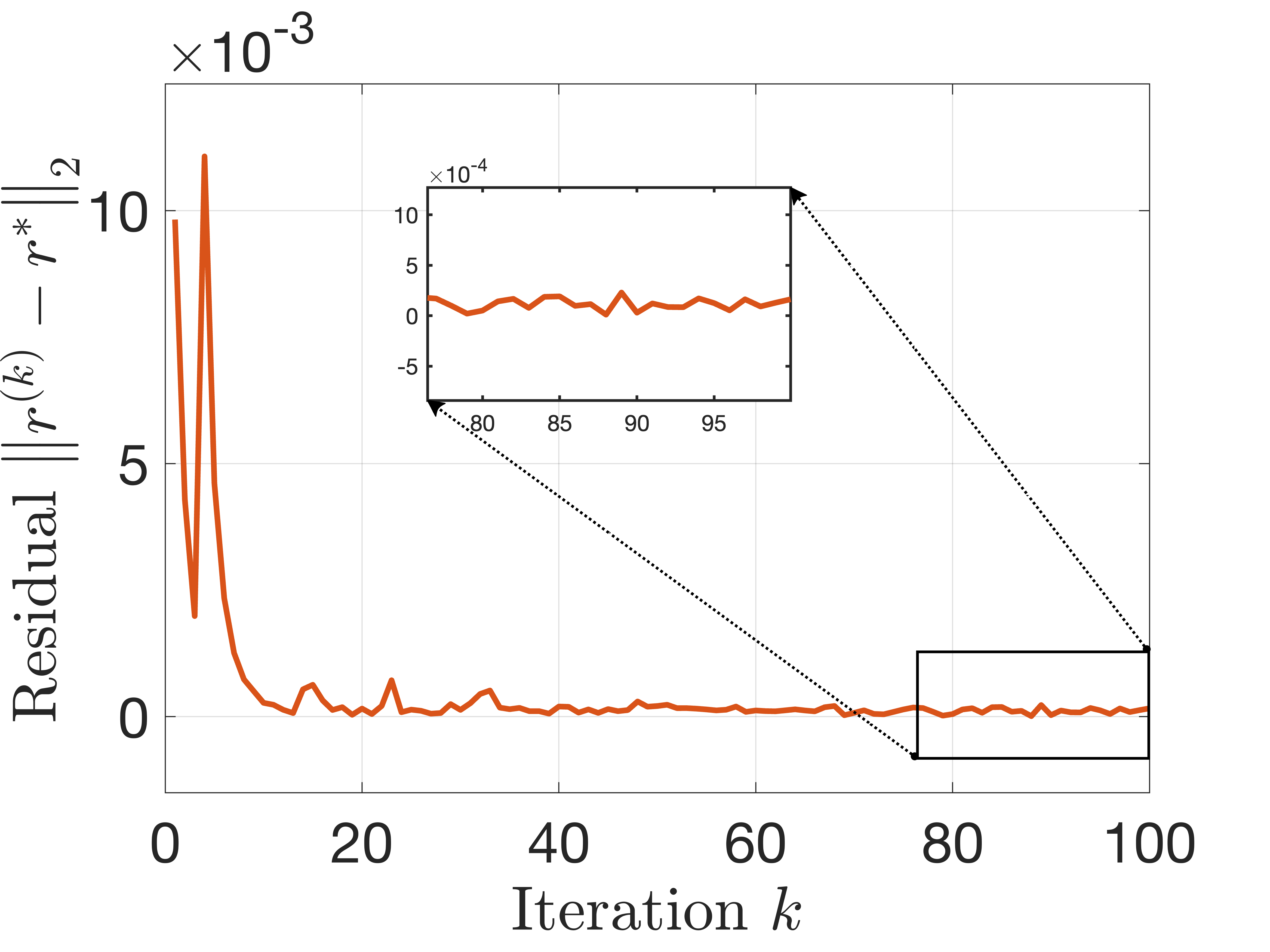}
	    \label{fig:3ENs_5APs} }
	    \hspace*{-2.15em} 
		 \subfigure[$N=3, M=50$]{
	     \includegraphics[width=0.245 \textwidth, height=0.135 \textheight]{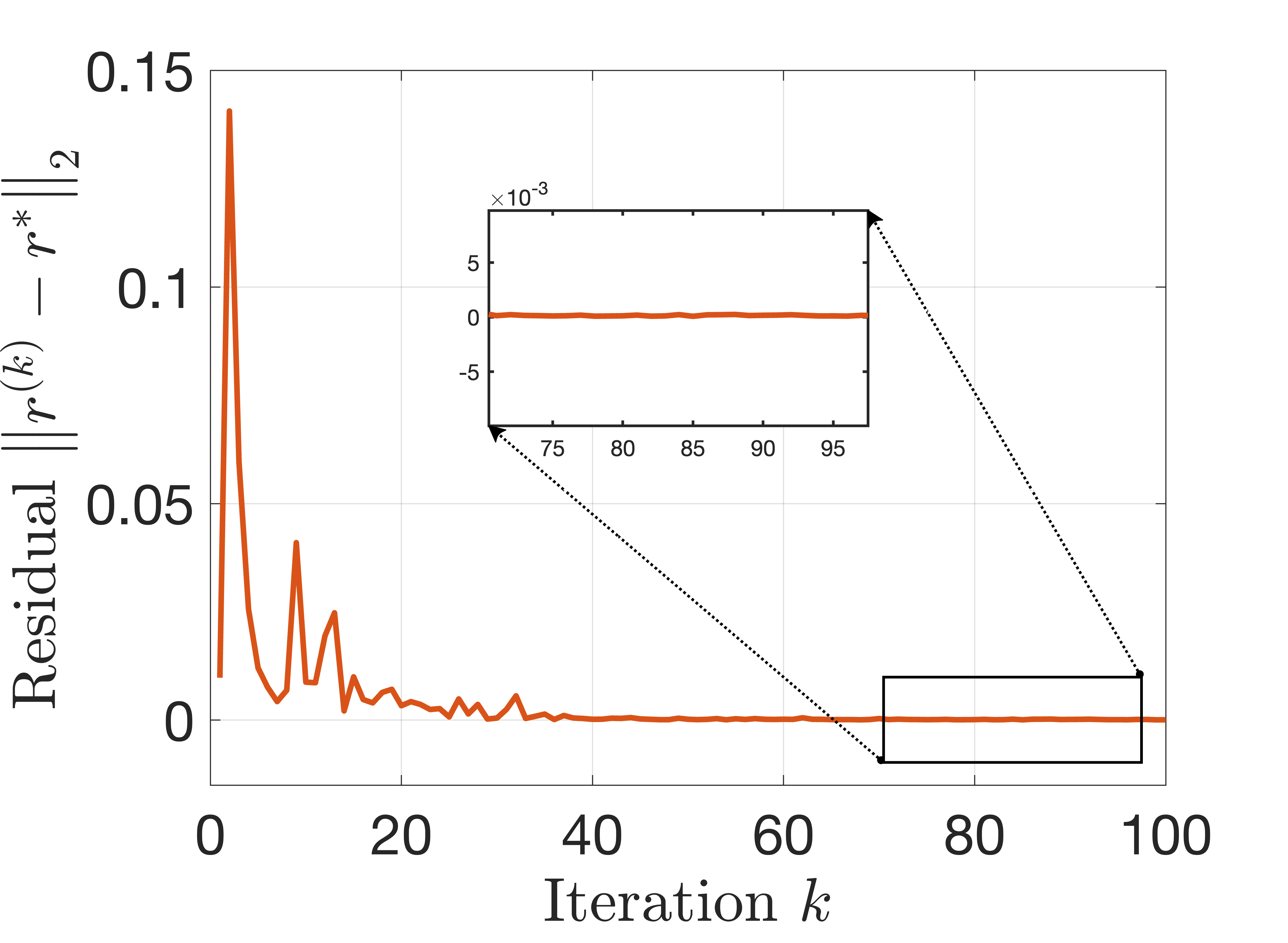}
	     \label{fig:3ENs_50APs}
	}  \vspace{-0.2cm}
	\caption{Convergence properties of the proposed algorithm.}
\label{fig:Convergence}
\end{figure}

\begin{figure}[ht!]
\centering
    \includegraphics[width=0.28\textwidth,height=0.145\textheight]{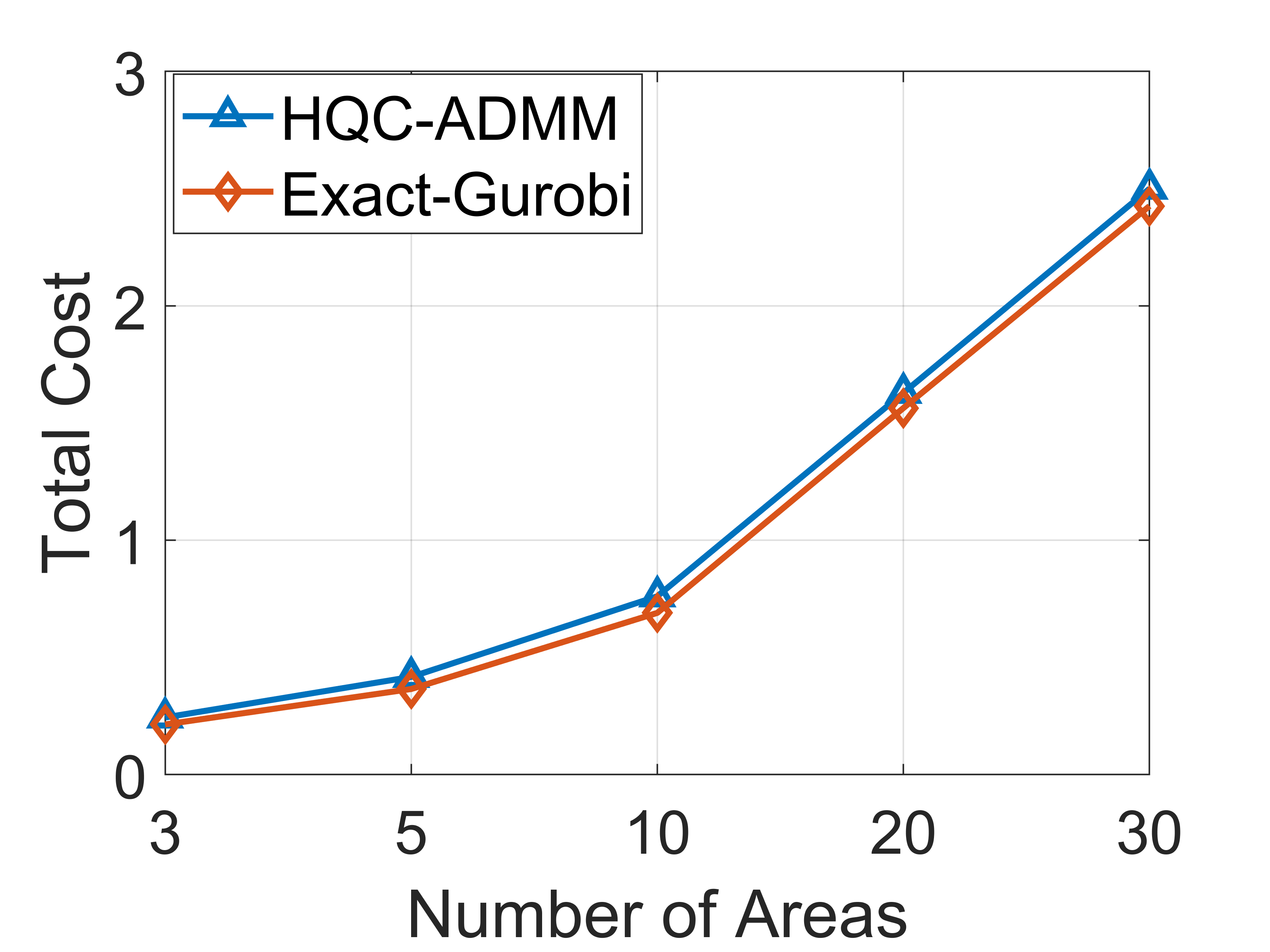}
	\caption{Performance comparison between the proposed  algorithm and the exact solution.}
\label{fig:Compare_Result}
\end{figure}

We first investigate the convergence property of the proposed algorithm, which is illustrated  in Fig. \ref{fig:Convergence} for different system sizes.
We can observe that the hybrid HQC-ADMM algorithm converges after a handful of iterations even when  the system size increases. Fig. \ref{fig:Compare_Result} shows that the total cost increases as the number of areas increases (i.e., increasing demand). More importantly, the results indicate that the solutions obtained by the proposed HQC-ADMM algorithm  are very close to the exact global optimal solutions obtained by the MILP solver (i.e., Gurobi). The accuracy of the proposed algorithm is further demonstrated in Table \ref{tab3}, which shows that both the HQC-ADMM algorithm and the MILP solver produce the same EN placement decision. 

Our experiments show that QC can be a promising approach for large-scale problems. Our simulation results confirm the theoretical analysis and demonstrate the fast convergence of hybrid quantum-classical ADMM algorithm in various settings. However, due to the heuristic nature, its results are not as accurate as those obtained by classical optimizers on classical computers. Therefore, the application of QC on critical operations of edge networks requires further investigation, considering the mathematical foundation behind quantum algorithms.

\begin{table}
    \centering
    \caption{Comparison of the optimal placement decision variables involved in different approaches.}
    \label{tab3}
    \setlength{\tabcolsep}{6pt}
    \begin{tabular}{|c|c|c|}
    \hline \rule{0pt}{12pt}
        Scenario&  Exact solution&    Proposed algorithm\\[3pt]
    \hline \rule{0pt}{10pt}
        $\left(3, 5 \right)$&  $y^*=\left(0, 0, 1 \right)^T$&    $\overline{y}^*=\left(0, 0, 1 \right)^T$\\[3pt]
    \hline \rule{0pt}{10pt}
        $\left(3, 50 \right)$&  $y^*=\left(1, 1, 1 \right)^T$&    $\overline{y}^*=\left(1, 1, 1 \right)^T$\\[3pt]
    \hline 
    \end{tabular}
\end{table}

\section{Conclusion and Future Work}
\label{conc}
In this paper, we proposed a hybrid quantum-classical  distributed algorithm by combining QC and ADMM methods to tackle the joint EN placement and workload allocation problem. The results show that, by adopting the integrated QC-ADMM algorithm, the platform can not only drastically improve the user experience but also effectively reduce its costs, while the run-time remains manageable. This implies that QC can be a promising approach for solving large-scale edge network design problems. However, the results are not as accurate as the results provided by classical optimizers on classical computers, because of the heuristic nature of QC solvers. Therefore, the application of QC on critical operations of EC requires more investigation considering the mathematical foundation behind quantum algorithms.

\bibliographystyle{IEEEtran}
\bibliography{Reference}



\end{document}